\documentclass[10pt,letterpaper]{article}
\usepackage[top=0.85in,left=2.75in,footskip=0.75in]{geometry}

\usepackage{amsmath,amssymb}

\usepackage{changepage}

\usepackage{textcomp,marvosym}

\usepackage{cite}

\usepackage[hidelinks]{hyperref}

\usepackage[nopatch=eqnum]{microtype}
\DisableLigatures[f]{encoding = *, family = * }

\usepackage[table]{xcolor}

\usepackage{array}

\usepackage{nth}

\newcolumntype{+}{!{\vrule width 2pt}}

\newlength\savedwidth

\newcommand\thickhline{\noalign{\global\savedwidth\arrayrulewidth\global\arrayrulewidth 2pt}%
\hline
\noalign{\global\arrayrulewidth\savedwidth}}

\raggedright
\usepackage[aboveskip=1pt,labelfont=bf,labelsep=period,justification=raggedright,singlelinecheck=off]{caption}

\makeatletter
\renewcommand{\@biblabel}[1]{\quad#1.}
\makeatother

\usepackage{lastpage,fancyhdr,graphicx}
\usepackage{epstopdf}
\fancyheadoffset[L]{2.25in}
\fancyfootoffset[L]{2.25in}
\usepackage{threeparttable}
\usepackage[utf8]{inputenc}
\usepackage{booktabs, caption, makecell}

\usepackage[symbol]{footmisc}
\usepackage{threeparttable}

\usepackage{geometry}
\usepackage{pdflscape}

\begin{document}
\vspace*{0.2in}

\begin{flushleft}
{\Large
\textbf\newline{Evidence for five types of fixation during a random saccade eye tracking task: Implications for the study of oculomotor fatigue} 
}
\newline
\\
Lee Friedman\textsuperscript{1,*},
Oleg V. Komogortsev\textsuperscript{1},
\\
\bigskip
\textbf{1} Department of Computer Science, Texas State University, San Marcos, Texas, USA
\\
\bigskip

* lfriedman10@gmail.com

\end{flushleft}
\section*{Abstract}
Our interest was to evaluate changes in fixation duration as a function of time-on-task (TOT) during a random saccade task.  We employed a large, publicly available dataset.  The frequency histogram of fixation durations was multimodal and modelled as a Gaussian mixture.  We found five fixation types.  The ``ideal'' response would be a single accurate saccade after each target movement, with a typical saccade latency of 200-250 msec, followed by a long fixation ($>$ 800 msec) until the next target jump.  We found fixations like this, but they comprised only 10\% of all fixations and were the first fixation after target movement only 23.4\% of the time.  More frequently (57.4\% of the time), the first fixation after target movement was short (117.7 msec mean) and was commonly followed by a corrective saccade. Across the entire 100 sec of the task, median total fixation duration decreased.  This decrease was approximated with a power law fit with $R^2=0.94$.  A detailed examination of the frequency of each of our five fixation types over time on task (TOT) revealed that the three shortest duration fixation types became more and more frequent with TOT whereas the two longest fixations became less and less frequent.  In all cases, the changes over TOT followed power law relationships, with $R^2$ values between 0.73 and 0.93.  We concluded that, over the 100 second duration of our task, long fixations are common in the first 15 to 22 seconds but become less common after that.  Short fixations are relatively uncommon in the first 15 to 22 seconds but become more and more common as the task progressed.  Apparently. the ability to produce an ideal response, although somewhat likely in the first 22 seconds, rapidly declines.  This might be related to a noted decline in saccade accuracy over time. 

\section*{Introduction}
We report on our observations regarding ocular fixation during a 100 second long random saccade task.  We employ a Gaussian mixture model to classify fixations into five fixation types based on fixation duration.  Next, we further characterize these fixation types along several dimensions.  Finally we evaluate the frequency of occurrence of each fixation type as a function of time on task (TOT).  Although we consider our results as related to the assessment of ocular fatigue, we note that all prior published assessments of ocular fatigue have employed much longer time intervals (from 18 min to 18 hrs, see \cite{Review}).  For this reason we will refer to our findings as related to TOT.  There is a substantial history of efforts to divide up fixations into fixation types based on fixation duration which we will briefly review.  Next, we will review prior findings relating fixation duration to either TOT or fatigue.

\subsection*{Prior literature on classification of fixation types based on duration.}
Saccade latencies are typically in the range of 200 msec \cite{LeighZee}.  There are two well known situations that produce very short saccade latencies.  First, fixations prior to corrective saccades are typically very short (100-130 msec) \cite{Becker,LeighZee}.  

The second type of short saccade latency is that prior to an ``express saccade'' \cite{LeighZee}. 

\begin{quote}
``Human subjects were asked to execute a saccade from a central fixation point to a peripheral target at the time of its onset. When the fixation point is turned off some time ($\approx$ 200 ms) before target onset, such that there is a gap where subjects see nothing, the distribution of their saccadic reaction times is bimodal with one narrow peak around 100 ms (express saccades) and another peak around 150 ms (regular saccades) measured from the onset of the target \cite{Express}.''
\end{quote}
When the initial fixation point is not turned off before the target appeared, this is referred to as an ``overlap'' trial.

Gezeck et al \cite{Gezeck} evaluated distributions of saccadic reaction times using a similar gap/overlap paradigm to \cite{Express}.  They noted three modes in histograms of fixation durations:  (1) an ``express'' mode (90-120 msec), (2) a ``fast regular'' mode (137-170 msec);  and (3) a ``slow regular'' mode (200-220 msec).  Multimodality was assessed using the excess-mass test \cite{excessmass}.  The multimodal histograms were fit using a Levenberg-Marquardt fit-procedure \cite{Press} assuming a superposition of multiple gamma distributions. 

Velichovsky \cite{Velichovsky} evaluated fixation duration distributions in the context of various visual memory tasks and noted that fixation duration distributions were ``strongly positively skewed".  He also noted that fixation duration distributions are not unimodal.  He reported a major mode near 180 msec and another mode near 100 msec that he labeled as "express fixations".

Nakatani and van Leeuwen \cite{Nakatani} evaluated fixations durations during a perceptual processing cognitive visual task and found evidence for 6 different fixation duration distributions.  Generally the best fitting distribution for all size fixation types was a logistic distribution.

Schleicher et al \cite{Schleicher} divided fixations into 3 categories based on duration:  (1) short fixations ($< 150~msec$, also labelled ``express'' fixations), (2) 150 - 900 msec (also labelled ``cognitive'' fixations), and (3) $>900~msec$ also labelled as ``overlong'' fixations.  

Galley et al \cite{Galley} divided fixation durations into 3 groups: (1) ``very short" ($<90~msec$), (2) express (90-150 msec) and (3) ``cognitive'' (150 - 900 msec).  Eye movements were measured using electrooculography (EOG).  They found that different cognitive tasks produced different patterns of fixations across these categories.

Velichkovsky et al \cite{eSports} studied eye-movements during a computer game, and compared low skilled amateurs, advanced amateurs and professional gamers.  They found that: 

\begin{quote}
  ``...in the low skill group the fixation duration distributions is highly uniform and unimodal. This suggests the presence of only one fixation cluster in this skill group. In the high skilled amateurs and, especially, in the professional players the distributions are bimodal with the two modes around 100 and 300 ms.  
\end{quote}

\noindent According to these authors, fixations in the range of 50-150 msec are ``ambient'' fixations and fixations in the range of 250-350 msec as ``focal'' fixations.  Ambient fixations do not permit conscious identification of objects whereas focal fixations reflect conscious perception of objects.

Similarly, Negi and Mitra \cite{Negi} evaluated the relationship between fixation duration and learning and divided fixation duration into three categories: (1) 66 msec to 150 msec (``ambient fixations''), (2) 300 to 500 msec (``focal'' fixations), and (3) fixation longer than 1000 msec (``too long'' fixations).  The presence of a large number of ``too long'' fixations was associated with reduced learning performance.  They propose that such long fixations may be an index of subjects either ``zoning out'' or being confused.

\subsubsection*{Fixation subtypes during reading}

Radach et al \cite{Radach} discussed three types of saccade latency that occur during reading.  When subjects finish reading a line and make a very large saccade moving to the beginning of the next line of text, they note that most (68\%) of these saccades are followed by corrective saccades with a latency of about 140 msec to 160 msec.  When these researchers remove the fixations prior to corrective saccades from the analysis, they find evidence of a population of "very short fixations" with a mean length typically shorter by at least 20 msec, than latencies prior to corrective saccades.  To account for these very short fixations they propose the "parallel programming hypothesis" which stipulates that there is a temporal overlap in the programming of successive saccades.  

Suppes \cite{Suppes} presents one of the earliest attempts to model fixation duration distributions during reading as a mixture of distributions.  Suppes proposed that fixation duration histograms could be best modelled as a mixture of an exponential distribution and a gamma distribution with shape parameter of 2.  According to McConkie and Dyre \cite{McConKie}, this model does not fit empirical fixation distributions well. 

McConkie and Dyre \cite{McConKie} point out that fixation duration histograms during reading have 3 periods:
\begin{quote}
``a slow increase in frequency up to about 150 ms, a sharp rise to a peak around
200 ms, and a long tail that reaches near-zero frequency around 500 ms.'' (page 684)
\end{quote}
They evaluate several competing reading models and conclude that the best model is a two-state transition model. After a forward saccade to a new word, subjects are in state 1, which lasts until the reader is able to start using the new information provided by the new word.  During this time, information from the previous fixation is still being processed by the CNS and the ability to acquire new information may be blocked.  During this state, the probability of a saccade is low.  State 2 begins when the system begins to acquire new information.  The probability of a saccade is higher in this state but its timing is modeled as random process.  Hazard functions are used to fit the two-state transition model.

According to Feng \cite{Feng}, Yang and McConkie \cite{Yang} provide a very strong case for three populations of fixations during reading:

\begin{quote}
``The strongest demonstration of multiple populations of fixations comes from a reading study by \cite{Yang}, in which they manipulated the information readers could perceive at any given fixation using the eye-movement -contingent display-change technique.  The manipulations of the text ranged from extreme (such as blanking the whole page) to modest (replacing text with non-words or filling all spaces with a symbol). Yang and McConkie found three distinct categories of fixations (Fig. 2 in \cite{Yang}).  The \textit{Early} fixations were short fixations (shorter than approximately 125 ms), which occurred regardless of experimental conditions. The \textit{Normal} fixations peaked at 175–200 ms. These fixations did not require linguistic information but the content being fixated needed to be ‘‘textlike.’’.  For instance, the distributions of these fixations were largely unaffected when a line of text was replaced by X`s with spaces preserved, but were severely suppressed when the spaces were removed. Lastly, the \textit{Late} fixations peaked roughly at around 350 ms and extended well beyond 700 ms in some cases. According to Yang and McConkie, these were results of cognitive inhibition in response to disruptions of visual information.'' \cite{Feng}, page 74.
\end{quote}

Feng \cite{Feng} modelled fixation duration histograms as mixtures of log-normal distributions and determined that three components provided the best fit.  The first two components were similar to the \textit{Early} and \textit{Normal} components described in Yang and McConkie \cite{Yang}, with the \textit{Early} peak at about 110 msec and the \textit{Normal} peak at 172 msec.  The \textit{Late} component in the Feng \cite{Feng} study peaked at 237 msec.

\subsection*{Prior literature relating fixation duration to fatigue or TOT.}
In Table 1 we review prior evidence linking fixation duration to either fatigue or TOT.  All of these studies lasted at least 18 minutes which is much longer than the duration (100 sec) of the random saccade task we employed in the present study.  Several studies looked at mean fixation duration and some studies evaluated different fixation types, based on fixation duration.  One study \cite{Fix2} found that mean fixation duration decreased with fatigue but two studies found that mean fixation duration increased with fatigue (\cite{Naeeri,Naeeri2}) and five studies found no relationship between mean fixation duration and fatigue (\cite{Schleicher,Cazzoli,Marandi,badminton,Pilot}).   Schleicher et al \cite{Schleicher} reported that ``express'' ($<150~msec$) fixations became more frequent with fatigue and ``cognitive'' fixations ($>150~msec$ and $<900~msec$) decreased with fatigue.  On the other hand, Zagari Mirandi et al. \cite{Marandi} reported that these same sort of fixations ($>150~msec$ and $<900~msec$) decreased with TOT.  Thus, no clear pattern emerges from the literature regarding the effect of fatigue on fixation duration

\subsection*{Present study}
The random saccade task we employed in the present study is unlike any prior task used in the study of eye-movements and fatigue.  The fact that every 1 sec a new target appeared means that our task is more structured than most prior tasks used in this research area.  This structure is likely to influence the fixation durations that we observe.  Our task is also much shorter (100 msec) than prior studies.  Also, our study was based on many more subjects (N=322) and fixations ($\approx$ 100,000) than other studies.  Therefore, our frequency distribution of fixation duration across subjects and repeat sessions has a unique, smooth and multimodal shape.  Following the approach of prior studies, we attempt to model our frequency distribution as a mixture of component distributions.  Unlike prior mixture distribution analyses, our histogram could be fit reasonably using Gaussian components.  As will become evident, our different fixation types each had a different but statistically significant relationships to TOT.

\newgeometry{margin=1cm} 
\begin{landscape}

\begin{table}[htbp]
\begin{adjustwidth}{0.8in}{0in}
\caption{Prior literature relating fixation duration to fatigue or time-on-task (TOT).}
\begin{threeparttable}
\begin{tabular}{|l|c|c|c|l|l|}
\hline
\textbf{First Author} & \textbf{Year} & \textbf{Cite} & \textbf{Task Duration} & \hspace{0.65in} \textbf{Task} & \textbf{Result} \\ \hline
Lavine  & 2002 & \cite{Fix2} & 30 min & visual pattern detection & Total fixation duration on visually displayed\\ 
& & & & & target digits decreased. \\ \hline
Schleicher & 2008 & \cite{Schleicher} & 134 min & self-rated alertness & Mean fixation duration unrelated to alertness. \\
& & & (average) & during simulated driving & Express fixations (\textless{}150 ms) become more frequent\\
& & & & &   with fatigue. \\ 
& & & & &  Cognitive fixations (150 - 900 msec) decrease with fatigue. \\ \hline
Cazzoli & 2014 & \cite{Cazzoli}  & $\approx$ 25 min & looking at urban and & No effect of time-on-task for fixation duration. \\ 
& & & & rural landscapes & \\ \hline
Naeeri  & 2018 & \cite{Naeeri} & 120 min & Flight simulator & Mean fixation duration increases over the 2 hours \\\
& & & & & for novice pilots but not experienced pilots.  \\ \hline
Zagari Marandi & 2018 & \cite{Marandi}  & 40 min & Memory and replication & No monotonic change in mean fixation duration.  \\ 
& & & &  of simple 2D pattern. &  Mean fixation duration  for fixations \textgreater 150 msec \\ 
& & & & & and \textless 900 msec increased with TOT. \\ \hline
Loiseau-Taupin & 2021 & \cite{badminton}  & \textgreater 18 min  & 3 phases: (1) play  & No effect of fatigue on mean fixation duration.  \\ 
& & & & badminton, (2) acute & \\ 
& & & & intense physical exercise & \\
& & & & (18 min), (3) play badminton & \\ \hline
Naeeri  & 2021 & \cite{Naeeri2}  & 240 min & flight simulation & Mean fixation duration increased with fatigue. \\ \hline
Lengenfelder & 2023 & \cite{Pilot}  & 70 min & computer image process- & No change in fixation duration from session 1 to session 2. \\
& & & & ing task - find moving vehicle & \\ \hline
\end{tabular}
\end{threeparttable}
\end{adjustwidth}
\end{table}

\end{landscape}
\restoregeometry

\section*{Materials and methods}
\subsection*{The Eye Tracking Database}
The eye tracking database employed in this study is fully described in \cite{GazeBase} and is labelled ``GazeBase''  It is publicly available (\url{https://figshare.com/articles/dataset/GazeBase_Data_Repository/12912257}).  All details regarding the overall design of the study, subject recruitment, tasks and stimuli descriptions, calibration efforts, and eye tracking equipment are presented there.  There were 9 temporally distinct "rounds" over a period of 37 months, and round 1 had the largest sample. This report only includes subjects from round 1. Briefly, subjects  were initially recruited from the undergraduate student population at Texas State University through email and targeted in-class announcements. A total of 322 subjects (151-F, 171-M) were included.  Subjects completed two sessions of recording (median 19 min. apart) for each round of collection. Each session consisted of multiple tasks. The only task employed in the present study was the random saccade task. During the random saccade task, subjects were to follow a white target on a dark screen as the target was displaced at random locations across the display monitor, ranging from ± 15 and ± 9 of degrees of visual angle (dva) in the horizontal and vertical directions, respectively. The minimum amplitude between adjacent target displacements was 2 dva. At each target location, the target was stationary for 1 sec.  There were 100 fixations per task (100,000 samples per task).  The target positions were randomized for each recording. The distribution of target locations was chosen to ensure uniform coverage across the display. Monocular (left) eye movements were captured at a 1,000 Hz sampling rate using an EyeLink 1000 eye tracker (SR Research, Ottawa, Ontario, Canada).

The gaze position signals for the random saccade task were classified into fixations, saccades, post-saccadic oscillations (PSOs) and various forms of artifact, using an updated version of our previously reported eye-movement classification method \cite{MNH}.

\subsection*{Screening Fixations:}
\subsubsection*{Removing fixations adjacent to artifact}
Any fixation which was immediately preceded or followed by any type of artifact (e.g., blink artifact), was excluded from this study. 

\subsubsection*{Removal of fixations that are part of Square Wave Jerks}
According to \cite{LeighZee}, one type of saccadic intrusion, ``Square Wave Jerks'' (SWJ)) are:
\begin{quote}
"...small (typically 0.5 degree), horizontal, involuntary saccades that take the eyes off the target and are followed, after an intersaccadic interval of about 250 milliseconds, by a corrective saccade that
brings the eyes back to the target. They may occur in normal individuals at frequencies of 20 per minute or greater." Page 250.
\end{quote}

Since fixations during these SWJ are fundamentally different from other fixation types, we wanted to exclude them.  To this end, we develop a MATLAB (Natick, Massachusetts) script to detect SWJ and remove the fixations associated with SWJ from our dataset.  We found a total of 1,467 SWJ.  Of 322 subjects, 54 had no SWJs.  Of the remaining 268 subjects, 55 had only 1 SWJ.  The median number of SWJ per subject was 2.5 (25th percentile = 1, 75th percentile= 5 SWJ per subject).  One subject had 26 SWJ in session 1 and 22 SWJ in session 2. Our MATLAB script for detecting SWJ, 61 eye movement datasets, and 89 example images of SWJ  are available online at \url{https://hdl.handle.net/10877/18499}.

\subsection*{Histogram of fixation lengths}
\begin{figure}[!ht]
\includegraphics[width=1.0\textwidth]{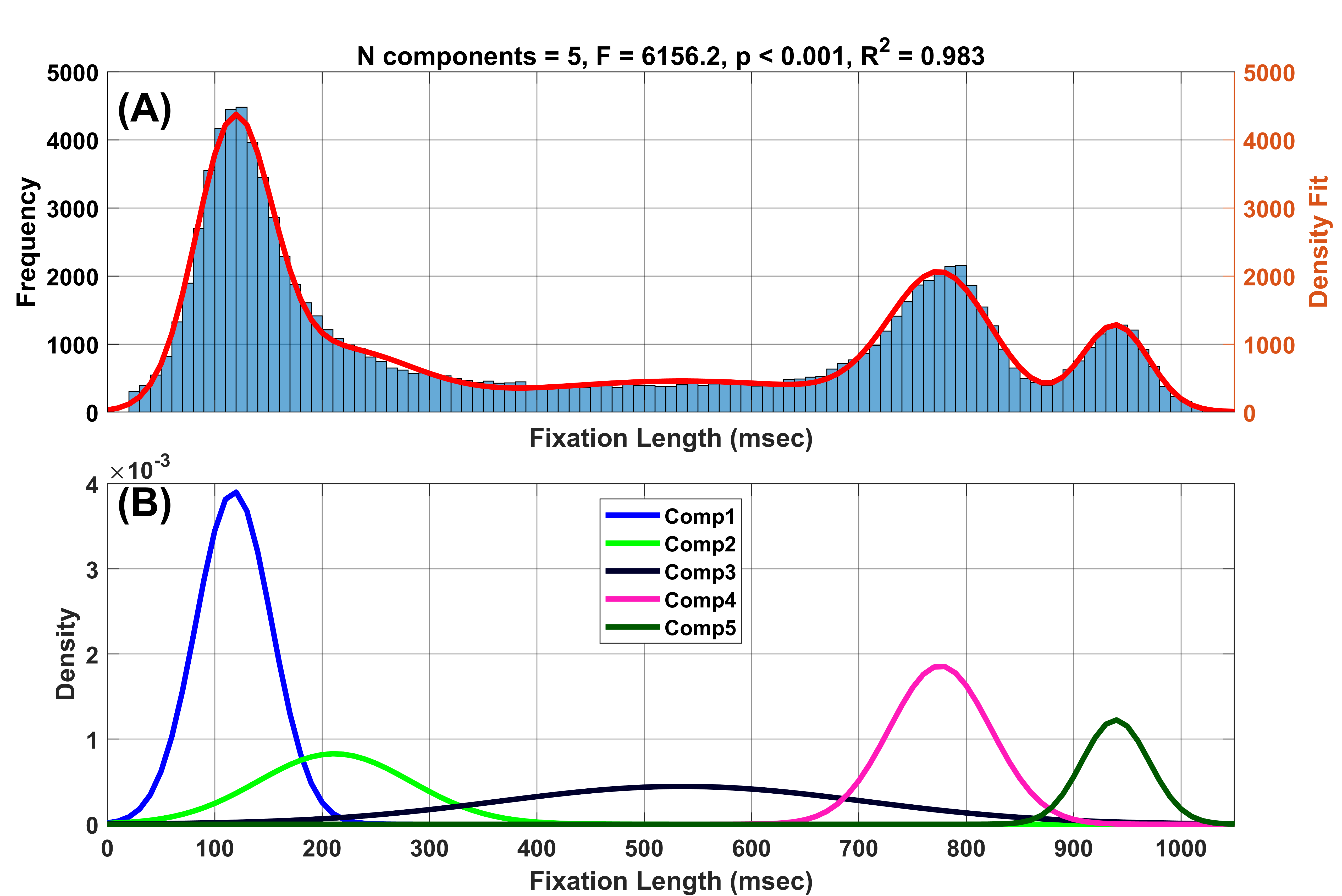}
\caption{{\bf Normal mixture distribution analysis of fixation length histogram.}  (A) Histogram of all fixation lengths (shown as blue bars).  The red curve is the fit of the five components illustrated in (B). (B) The five component distributions found.}
\label{fig1}
\end{figure}

From 322 subjects recorded on two sessions, we obtained 102,115 fixations.  The first step in the analysis was to create a frequency distribution of fixation lengths in ms (Fig. \ref{fig1}(A)).  This distribution was multimodal in appearance.  To further characterize the multimodality of this distribution, the frequency distribution was processed using a Gaussian mixture model analysis described below.

\subsection*{Gaussian Mixture Model Analysis}
\label{sec:MixtureModel}
We employed the mclust R package \cite{mclust} to fit from two to ten components to the histogram data shown in Fig. \ref{fig1} (A). For this analysis, the variances of each component were allowed to be unequal.  The R script for this is available at (ref: http.www.digital.collections.txstate).  Each normal component is represented by a mean, a standard deviation (SD), and a weight. The sum of these weights is always 1. 

\begin{figure}[!ht]
\includegraphics[width=1.0\textwidth]{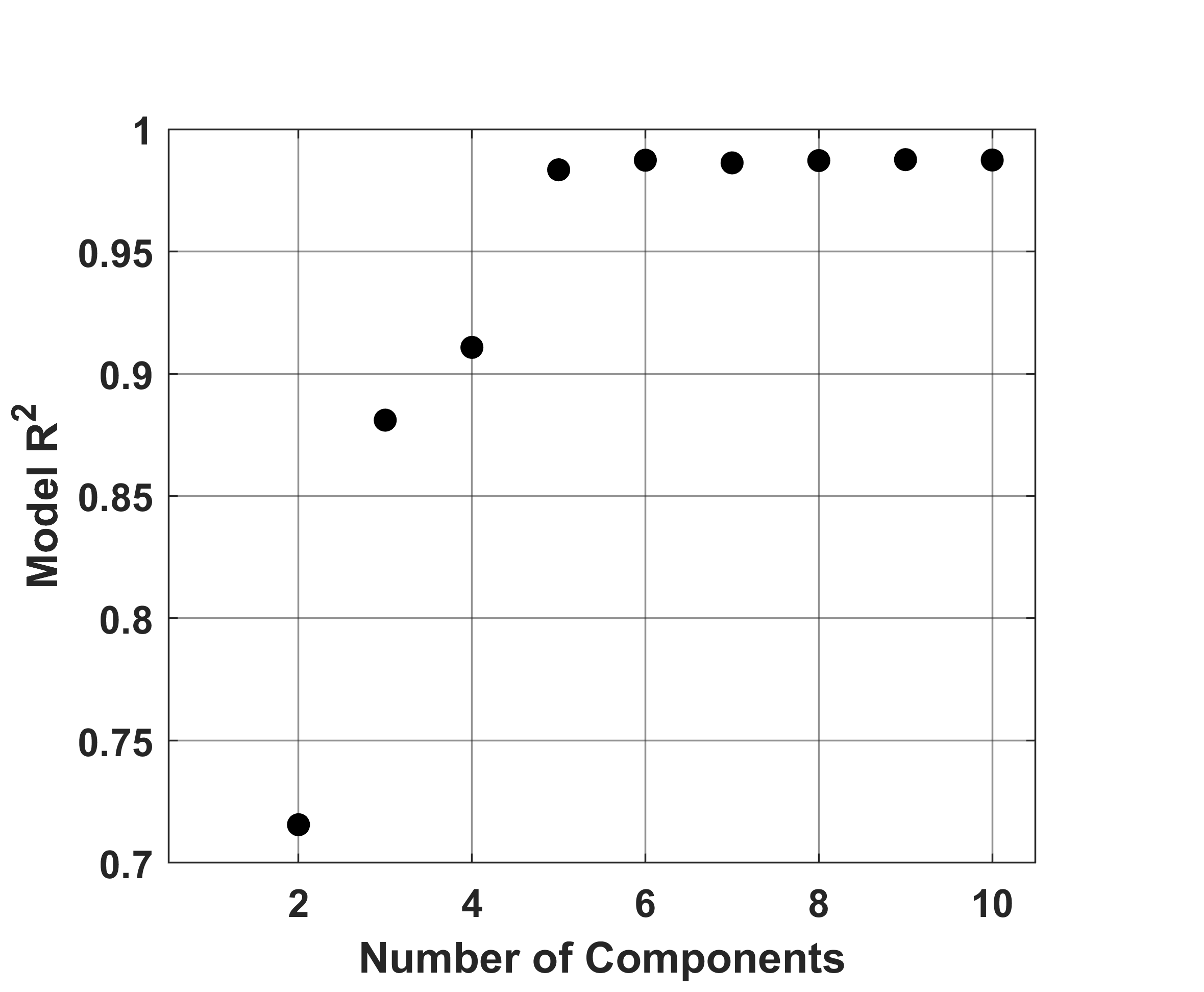}
\caption{{\bf Model $R^2$s for each  number of components (2:10) resulting from each mclust model.}}  
\label{fig2}
\end{figure}

To determine which of the nine fits was best, we computed a density curve, using the means, SDs and weights for each analysis evaluated at the same intervals as the histogram in Fig. \ref{fig1} (A).  We then regressed these density curves onto the distribution of counts from the histogram.  The results of this fit for five normal components is shown as a red curved line in Fig. \ref{fig1} (A).  We computed the model $R^2$ for each of the nine density curves based on  number of components and plotted these model $R^2$s as in Fig. \ref{fig2}.  It appears from Fig. \ref{fig2} that almost nothing is gained in the degree of fit after 5 components have been fit.  Therefore, we consider that there are five different fixation types. 
The weights, means and SDs for each of the five components are presented in Table 2.

\begin{table*}
\centering\begin{minipage}{\textwidth}
\caption{\textbf{Weights, Means, SDs and Limits for each Fixation Type.}}
 \begin{tabular}{|c|c|c|c|c|c|}
 \thickhline
  Fixation Type: & 1 & 2 & 3 & 4 & 5 \\  \hline
  Weight \footnote{Weights can be interpreted as proportions of all fixations of each type.} & 0.35  & 0.15  & 0.19  & 0.22  & 0.1 \\ \hline
  Means (ms) & 117.7 & 211.4 & 534.9 & 775.7 & 939.1 \\ \hline
  SDs (ms) & 35.3  & 71.4  & 170.4 & 47.2  & 31.1  \\ \hline
  Min (ms) \footnote{Defined by membership probability.}  & 21  & 182 & 329  & 686 & 878 \\ \hline
  Max (ms) \textsuperscript{\textit{b}}  & 181 & 328 & 685  & 877 & 1041  \\ \hline
  Means(P) \textsuperscript{\textit{b}} & 117.3 & 239.5 & 515.1 & 775.7 & 938.6\\  \hline
  \end{tabular}
\end{minipage}
\end{table*}

\subsection*{Assigning fixations to fixation types.} 
\begin{figure}[!ht]
\includegraphics[width=1.0\textwidth]{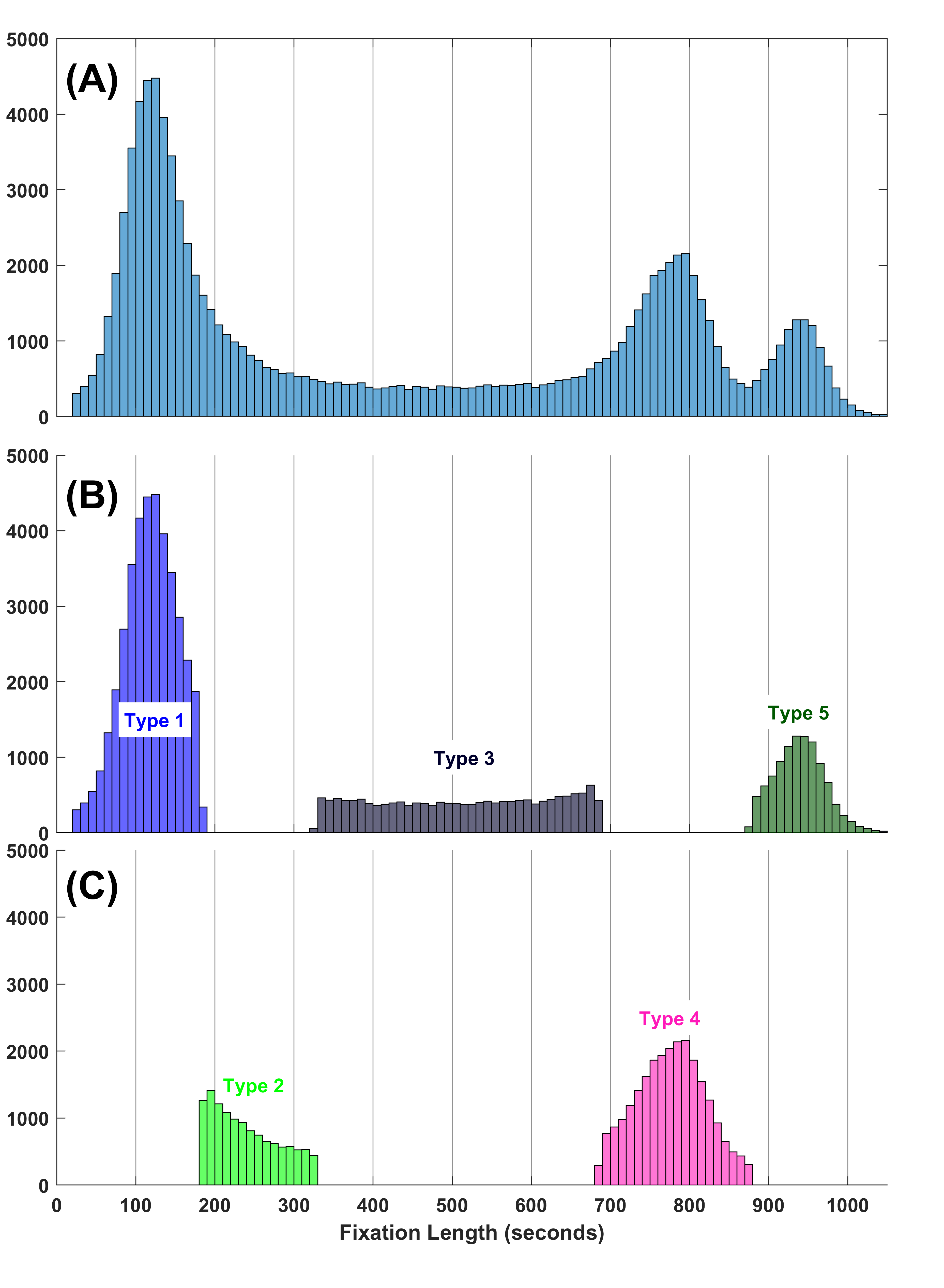}
\caption{{\bf Probability-based assignment of fixation lengths to one of the five fixation types.} (A) The histogram of fixation lengths. (B) and (C) Fixations divided up into fixation types.}  
\label{fig3}
\end{figure}

Theoretically, each fixation has some probability of belonging to each of the five component distributions.  The mclust R package \cite{mclust} assigns each fixation to a fixation type based on the maximum probability that each fixation is a member of each of the five components. This assignment in illustrated in Fig. \ref{fig3}.  From this analysis, we were able to divide up all of our fixations into one of five types based on the minima and maxima fixation length for each component.  The minima and maxima as well as the means for each group based on the probability assignment are also presented in the last three rows of Table 2.

\subsection*{Finding which Fixation Types are Followed by Corrective Saccades}
We were interested in determining which fixation types were followed by corrective saccades.  Since a target step might induce a saccade we only checked for corrective saccades at the end of fixations during which there was no target change.  To be classified as a corrective saccade, the last sample of the saccade had to reduce Euclidean distance between the prior fixation and the target. 

\subsection*{Fitting power law functions to assess time-on-task effects}
We wanted to evaluate the frequency of occurrence of each Fixation Type over time within each 100,000 sample task.  We ignored the first sec of data and the last sec of data because we thought these time periods might not be representative.  We divided the task into 14 seven sec intervals starting from 1 to 99 sec.  So the intervals started at  1, 8, 15, 22, 29, 36, 43, 50, 57, 64, 71, 78, 85, and 92 sec.  We counted the number of each Fixation Type that occurred within each time interval.  For each Fixation Type, we fit a power law function ($a*x^b$) using MATLAB's (Natick, MA) curve fitting application.  These fits produced estimates of $a$ and $b$ as well as an adjusted $R^2$.  These power law functions can be linearized by plotting frequency vs log(time interval).  These linearized forms can be assessed with linear regression and an $F$, $df$ and \textit{p-value} can be obtained.

\subsection*{Saccade Accuracy}
We were interested in determining if saccade accuracy declined over time.  Only the first saccade after each target movement was evaluated.  To be included in the in the analysis, the following criteria had to be met: (1) saccade latency was between 150 and 350 msec; (2) there was no artifact  from the time of the target change to the end of the saccade plus 100 msec.  The Euclidean distance between the eye position at the end of the saccade and the target was taken as the saccade error.  We looked at mean saccade error as a function of time block.  We also computed the ``percent saccade error'' by dividing each saccade error measurement by the size (Euclidean distance) of the prior target movement.  This was also related to time blocks.

\subsection*{Statistical Analysis}
We determined the proportion of the time that a target step is followed by each one of our fixations.  We also determine the proportions of fixation types that were followed by a corrective eye movement.  For each analysis, the proportions were compared with using a Tukey's HSD multiple comparisons procedure 
\cite{tmcomptest}.  

We also assessed the time between the preceding target change and the subsequent fixation start for each fixation type.  We tested for Fixation Type differences using the Kruskal-Wallis test which produces a $\tilde{\chi}^2$, $df$ and \textit{p-value}. A statistically significant Kruskal-Wallis test was followed up with a multiple comparison test which controlled for multiple comparisons with a Tukey HSD procedure ($\alpha = 0.05$).

\section*{Results}
\subsection*{Histogram of fixation lengths and results of mixture distribution analysis}
Fig. \ref{fig1} (A) presents the histogram of all 102,115 fixations from 322 subjects, each with 2 recording sessions.  The average fixation length across all fixations was 434.4 ms (SD: 318.2 ms).  On average, each subject had 159.4 (SD=35.0) fixations per session.  The median length of a fixation was 315 samples (\nth{25}, \nth{75} percentiles = 132, 763).

Fixation Type 1 (the briefest fixation) was the most numerous (35\% of all fixations), whereas the longest fixations (Type 5) were relatively rare (Table 2).  The fit of this model to the raw histogram is illustrated with a red curve in Fig. \ref{fig1} (A).

\subsection*{Exemplars of each fixation type.}
\begin{figure}[!ht]
\includegraphics[width=1.0\textwidth]{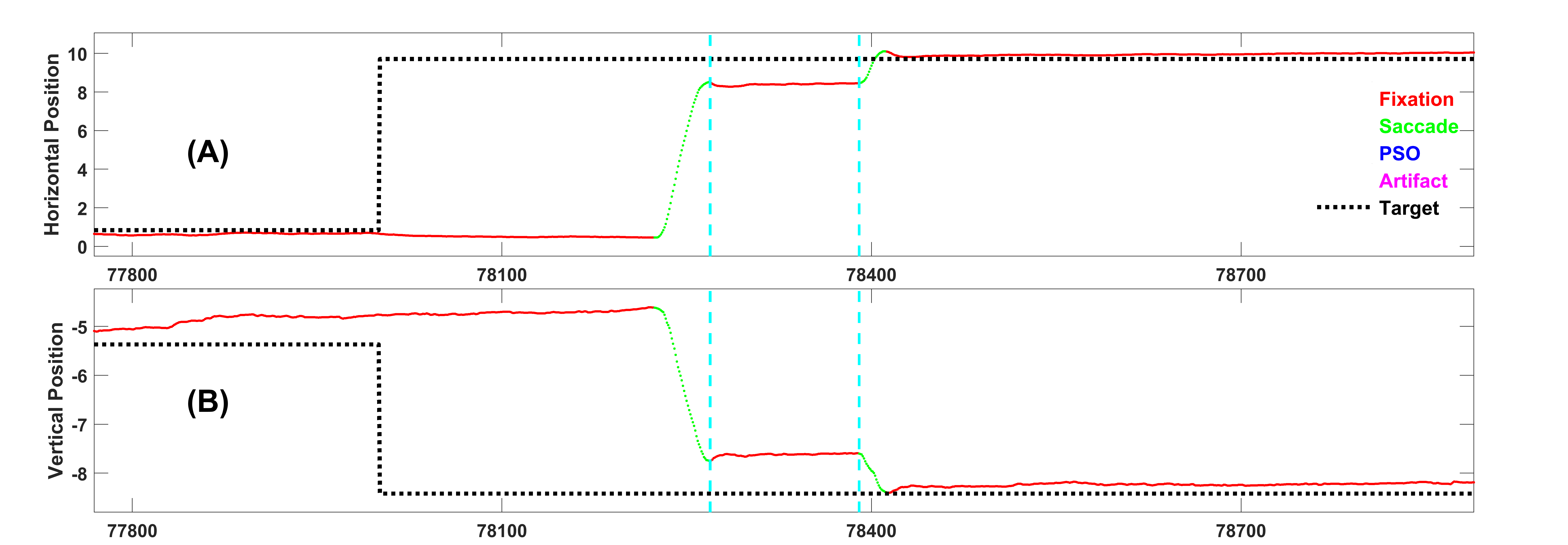}
\caption{{\bf Representative example of Fixation Type 1.} This fixation is 121 samples long and occurs 268 ms after the target changed position.  Note the corrective saccade following this short fixation. This is typical of this type of fixation.}
\label{fig4}
\end{figure}

\begin{figure}[!ht]
\includegraphics[width=1.0\textwidth]{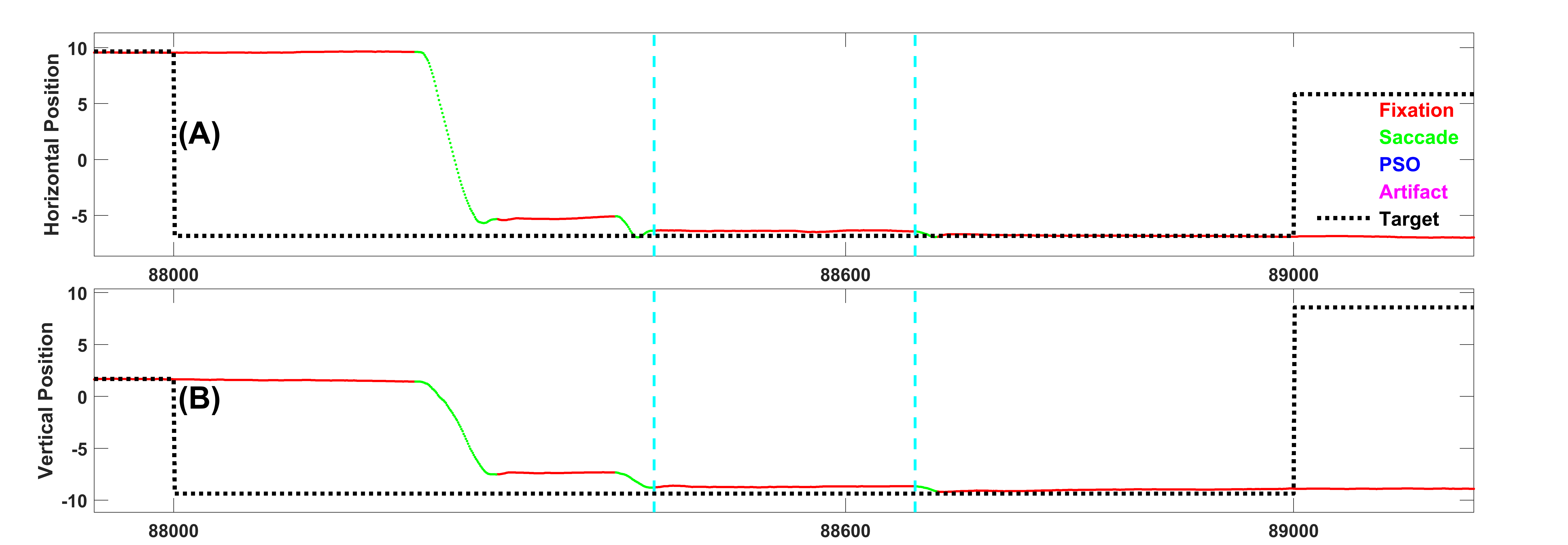}
\caption{{\bf Representative example of Fixation Type 2.} This fixation is 233 samples long and occurs 428 ms after the target changed position.  It is preceded by a Fixation Type 1.}
\label{fig5}
\end{figure}

\begin{figure}[!ht]
\includegraphics[width=1.0\textwidth]{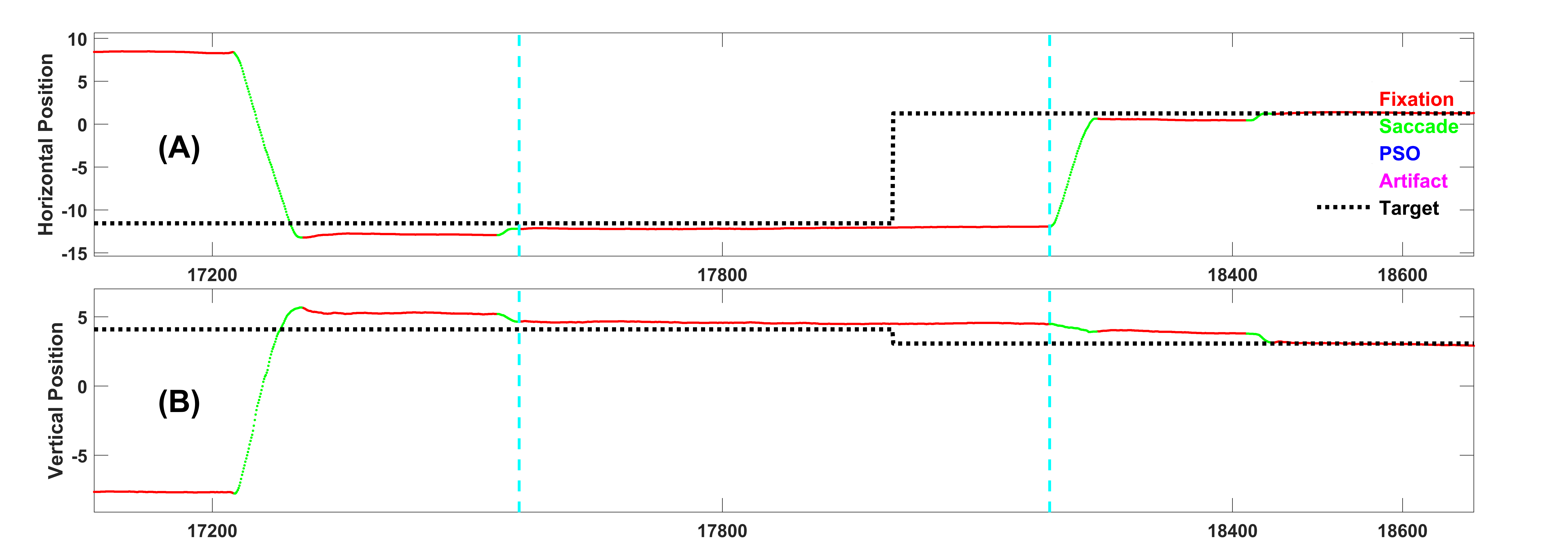}
\caption{{\bf Representative example of Fixation Type 3.} This fixation is 624 samples long and occurs 560 ms after the target changed position.}
\label{fig6}
\end{figure}

\begin{figure}[!ht]
\includegraphics[width=1.0\textwidth]{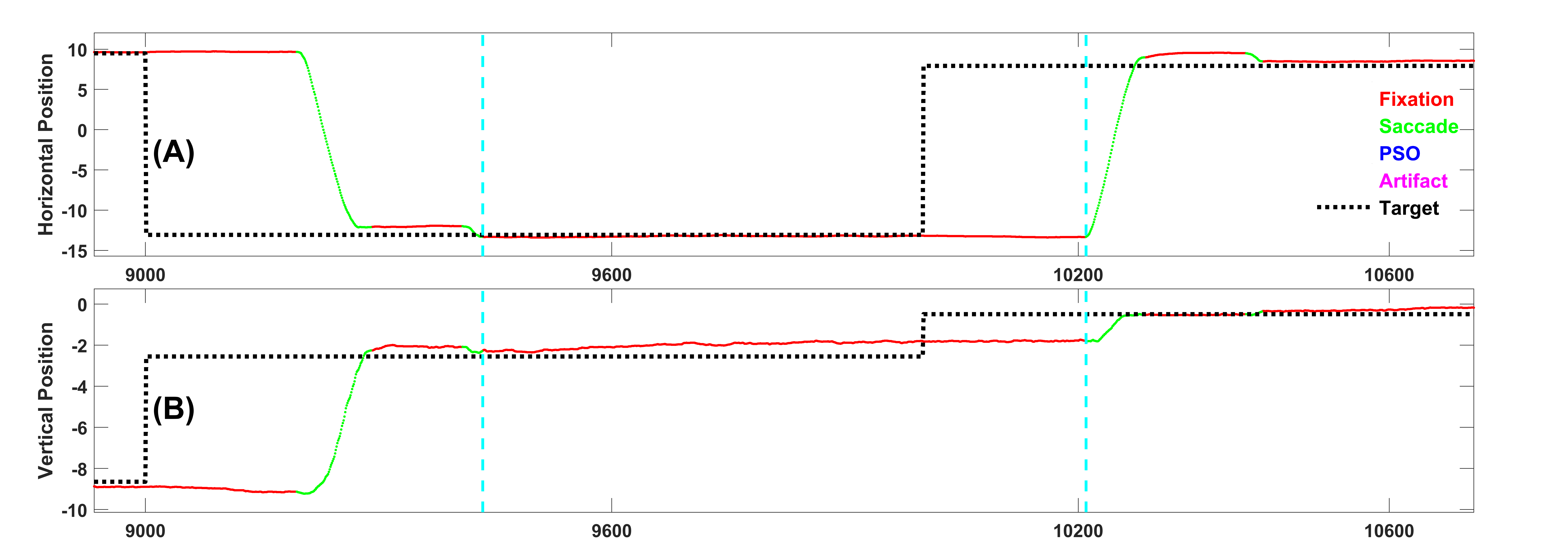}
\caption{{\bf Representative example of Fixation Type 4.} This fixation is 776 samples long and occurs 433 ms after the target changed position.  It is preceded by a Fixation Type 1.}
\label{fig7}
\end{figure}

\begin{figure}[!ht]
\includegraphics[width=1.0\textwidth]{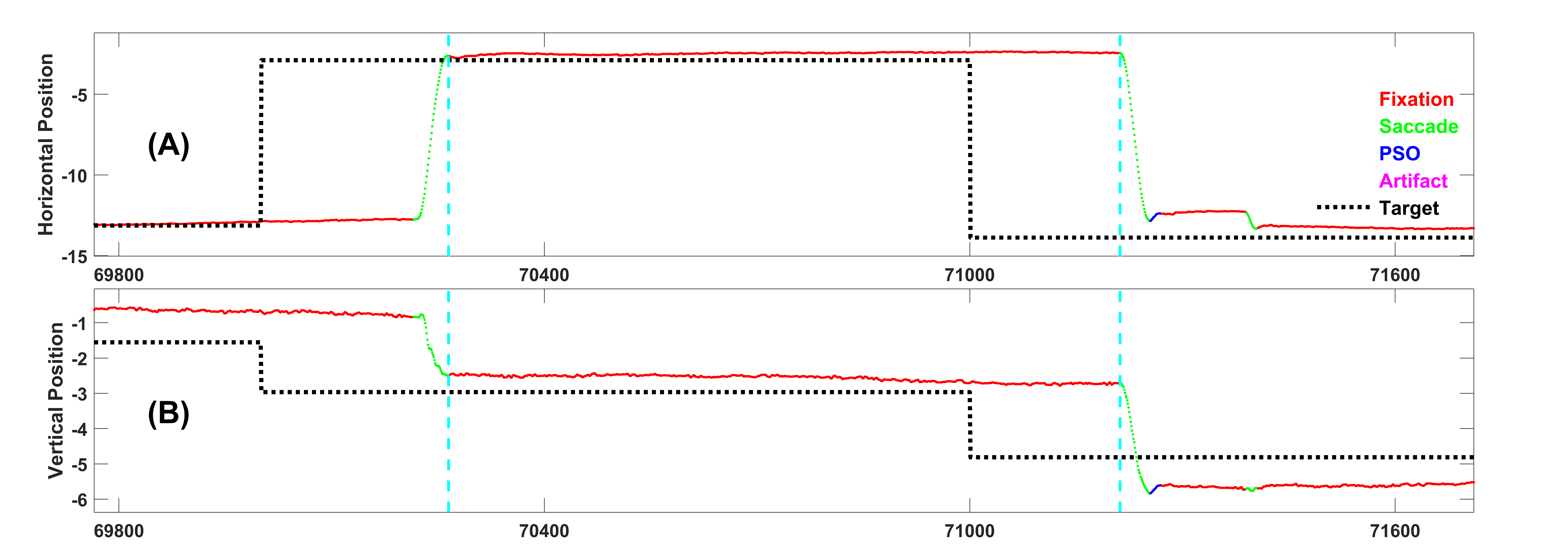}
\caption{{\bf Representative example of Fixation Type 5.} This fixation is 947 samples long, and occurs 264 ms after the target changed position.  Note that this is the ``ideal'' response, i.e., the subject sees the target step and makes one accurate saccade to the target and does not move until the next target step. Although this is the ideal response, as we will see below, it is relatively rare.}
\label{fig8}
\end{figure}

Figs. \ref{fig4},\ref{fig5},\ref{fig6},\ref{fig7},and \ref{fig8} each present one representative fixation of each fixation type.

\subsection*{Characterizing the five fixation types.}
\subsubsection*{Which Fixation Types occur first after target steps.}
\begin{figure}[!ht]
\includegraphics[width=1.0\textwidth]{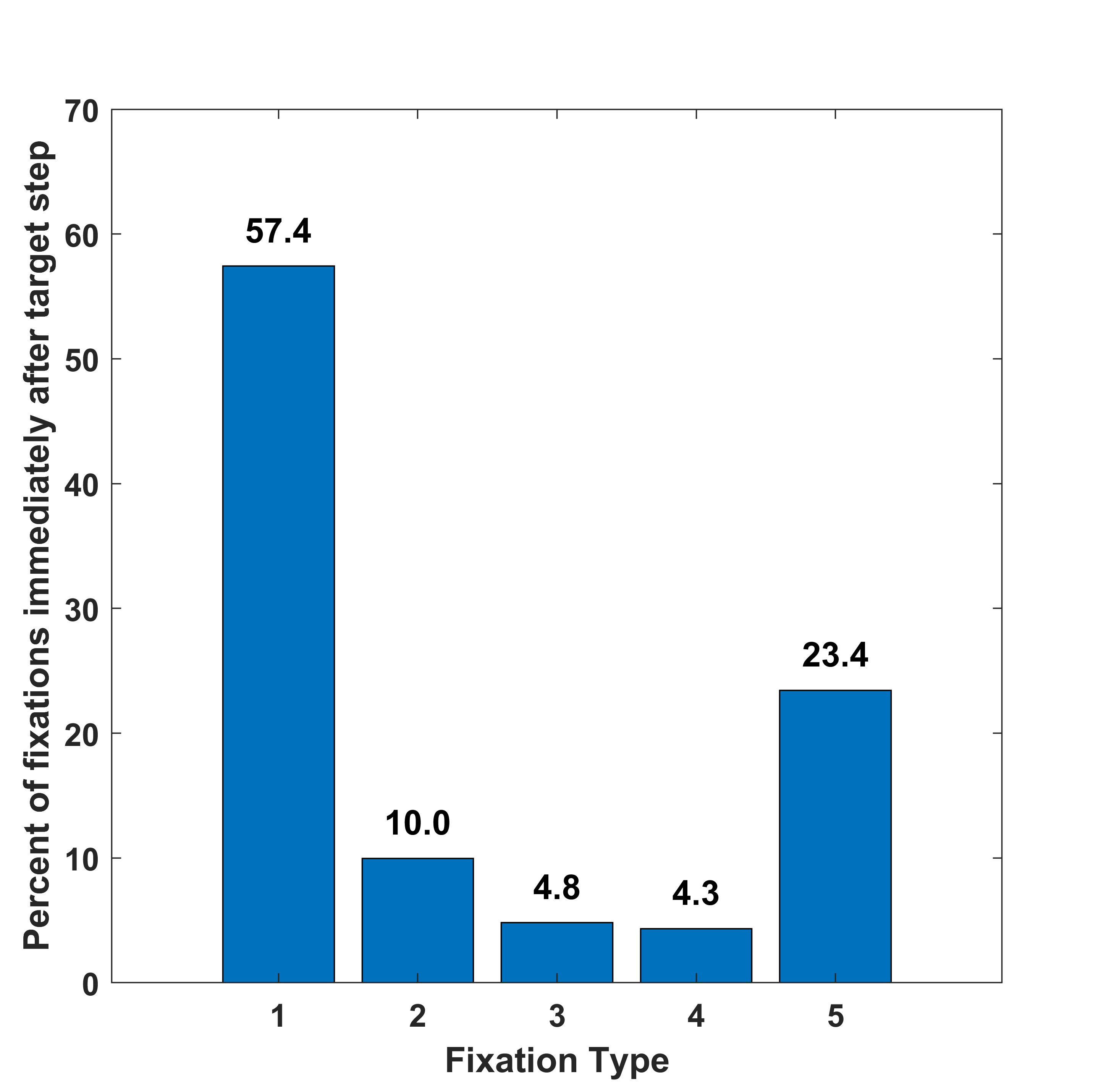}
\caption{{\bf Bar chart of the percentage of time that a target step is immediately followed by each one of the Fixation Types.}}
\label{fig9}
\end{figure}

In Fig. \ref{fig9} we present the percentage of fixation types that immediately follow target steps.  These values are based on an analysis of 32,935 target steps. The modal saccade latency across the dataset was 198, the median latency was 204 ms (\nth{10}, \nth{90} percentiles: 177 ms, 253ms).  To exclude anticipatory saccades which start before the target moves, target steps that were followed by a saccade with a latency less than 150 ms (anticipatory saccades) were excluded. Approximately 60\% of the time, a target step was followed by a Fixation Type 1, the shortest type (mean $\approx$ 120 ms). Fixation Types 3 and 4 occur very rarely after a target step.  The longest Fixation Type, Type 5, occurs $\approx$ 23\% of the time.  Fixations Types 1 and 5 were the first fixation type $\approx$ 80\% of the time.  Each of these five proportions were statistical significantly different from all others.  

\subsection*{Which Fixation Types are Followed by a Corrective Saccade?}
\begin{figure}[!ht]
\includegraphics[width=1.0\textwidth]{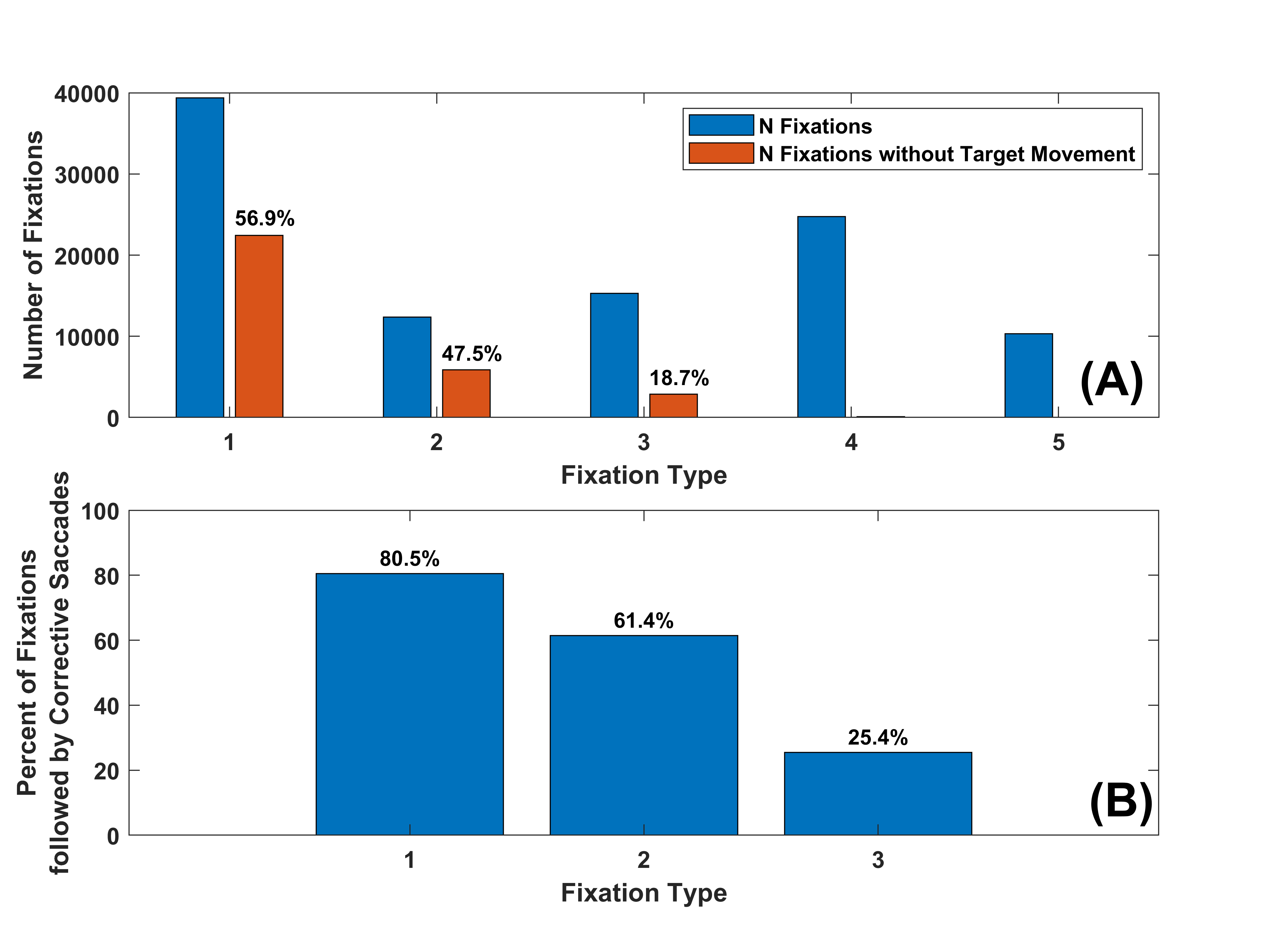}
\caption{{\bf Bar chart of the percentage of time that a target step is followed by a corrective eye movement.} (A) The blue bars represent the number of total fixations of each type.  The red bars are the number of each fixations of each type during which there was no target change.  (B) These are the percentages of fixations during which the target did not change that are followed by a corrective saccade.}
\label{fig10}
\end{figure}

We wanted to know how often each fixation type was followed by a corrective saccadic movement toward the target.  The results are presented in  Fig. \ref{fig10}.  For $\approx$ 57\% of the time there was no target step during Fixation Type 1 events (Fig \ref{fig10} (A)).  For $\approx$ 48\% of the time there was no target step during Fixation Type 2 events (Fig \ref{fig10} (A)).  For only $\approx$ 19\% of the time there was no target step during Fixation Type 3 events (Fig \ref{fig10} (A)).  There were very few Fixation Type 4 or 5 events that were not interrupted by a target step.  

For $\approx$ 81\% of the time that the target did not move during Fixation Type 1, the following saccade was corrective (Fig \ref{fig10} (B)).  For $\approx$ 61\% of the time that the target did not move during Fixation Type 2, the following saccade was also corrective (Fig \ref{fig10} (B)).  Fixation Type 3 events without an intervening target change were rarely followed by a corrective saccade (Fig \ref{fig10} (B)).

\subsubsection*{Time between preceding target change and the subsequent fixation start} 
\begin{figure}[!ht]
\includegraphics[width=1.0\textwidth]{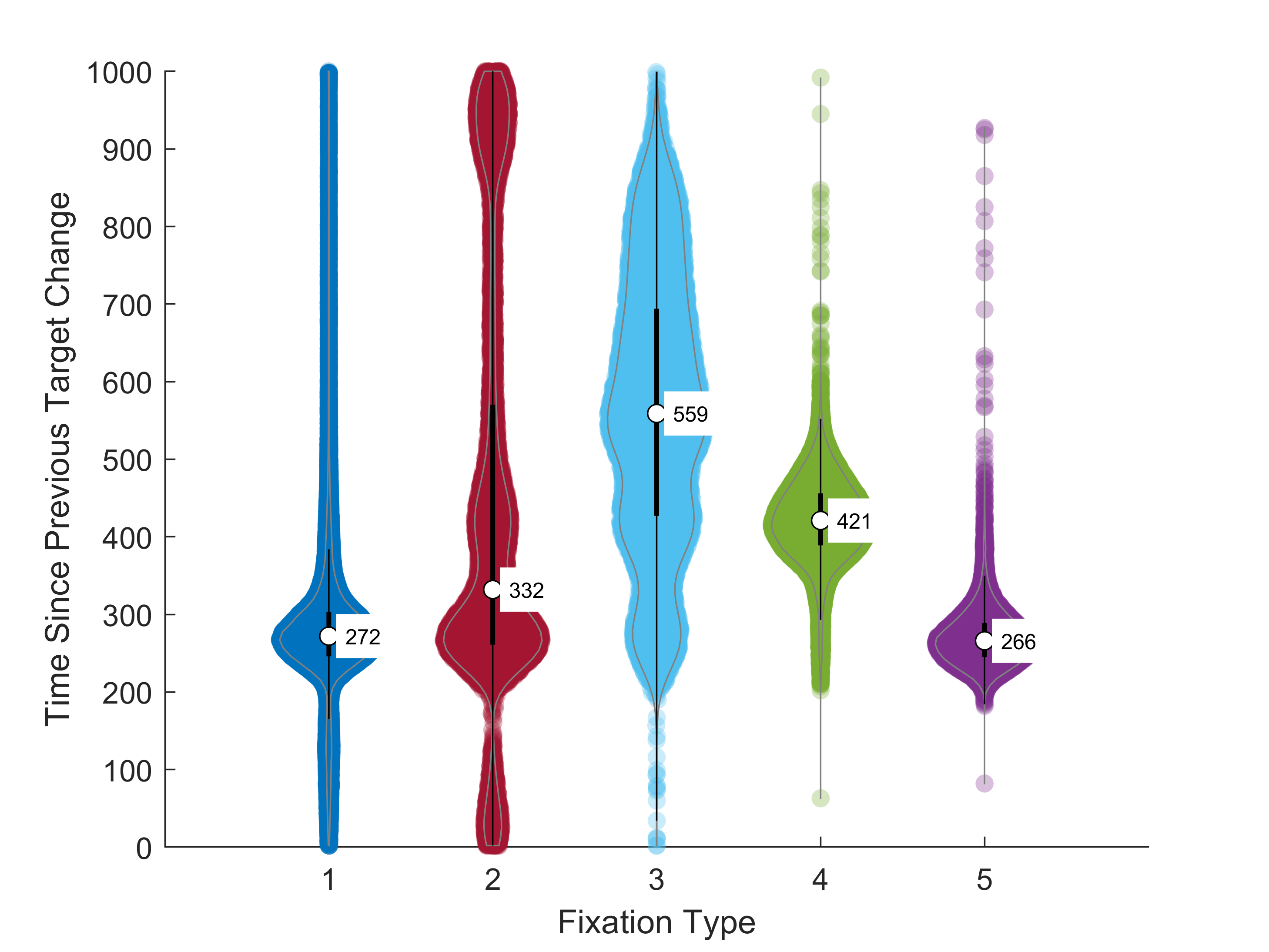}
\caption{{\bf Time between preceding target change and the start of the each Fixation Type.  The numbers are medians.}}
\label{fig11}
\end{figure}

In Fig. \ref{fig11} we present the distributions (as violin charts) of the time between the preceding target change and the beginning of each fixation for each fixation type.  The  $\tilde{\chi}^2$ from the Kruskal-Wallis test was 48,190.5 $(df = 4, 99498), p < 0.0001$).  Post-hoc multiple comparisons indicated that on this metric, each fixation type was statistically significantly different all others.  Fixation types 1 (the briefest) and fixation type 5 (the longest) tend to occur soon after target change.  The other fixation types, especially Types 4 and especially 3, tend to start later.

\subsection*{Change in Median Fixation Duration over TOT} 
\begin{figure}[!ht]
\includegraphics[width=1.0\textwidth]{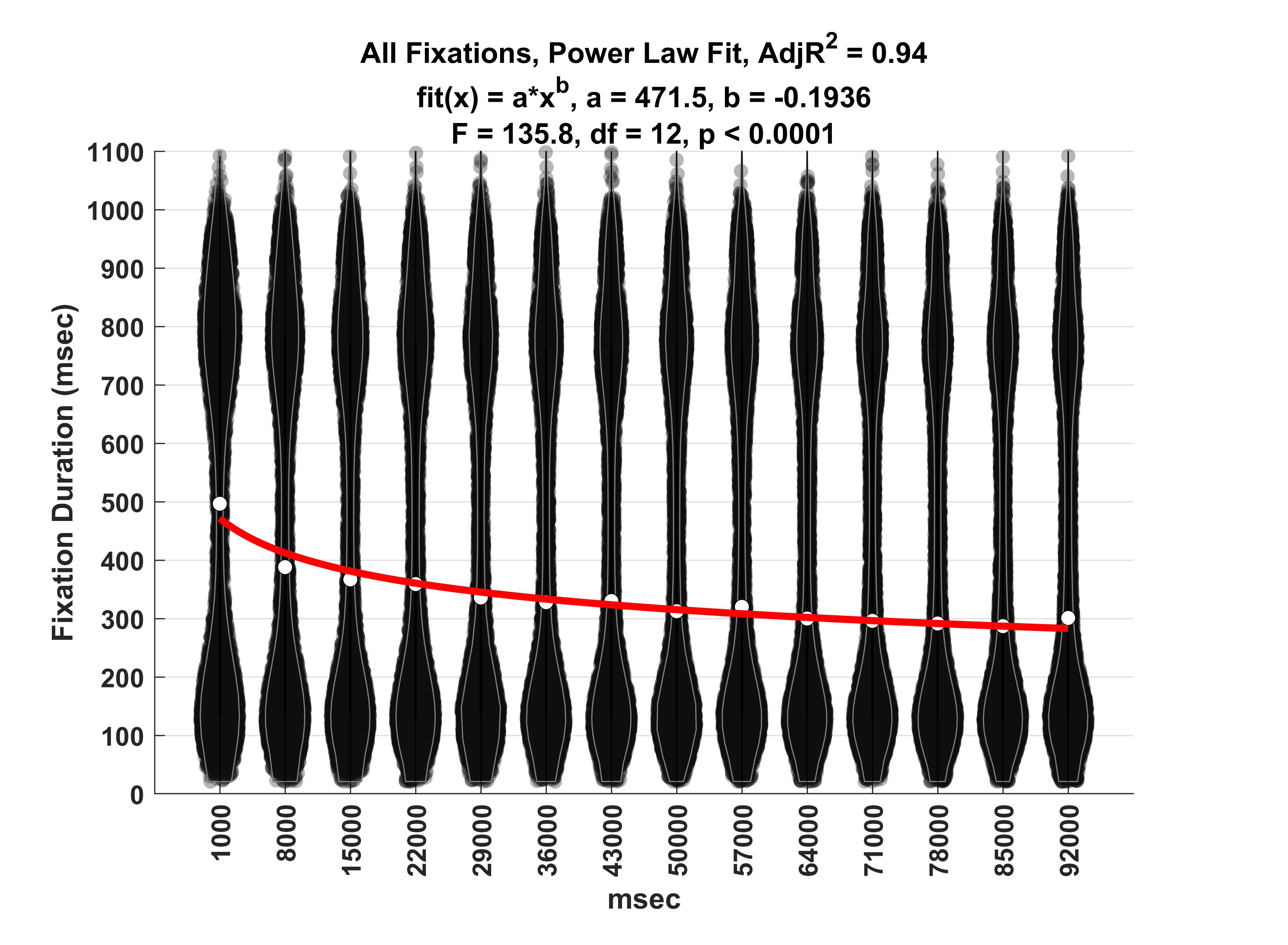}
\caption{{\bf Plot of the mean fixation duration change over TOT.} White dots represent medians.  Red line is the power law fit to the medians.}
\label{fig12}
\end{figure}

Figs \ref{fig12} is a set of violin plots illustrating the decrease in median fixation duration over TOT.  As noted above, time periods start at 1 sec into the task, and each time period includes 7 sec of data.  The seven sec time periods analyzed start at 1 sec (1, 8, 15, 22, 29,  36, 43, 50, 57, 64, 71, 78, 85 and 92).  Note the curvilinear decline in median fixation duration over time.  The power law relationship was statistically significant and accounted for 94\% of the variance of the medians.  

\subsection*{Analysis of the number of fixations of each type over time.} 
\begin{figure}[!ht]
\includegraphics[width=1.0\textwidth]{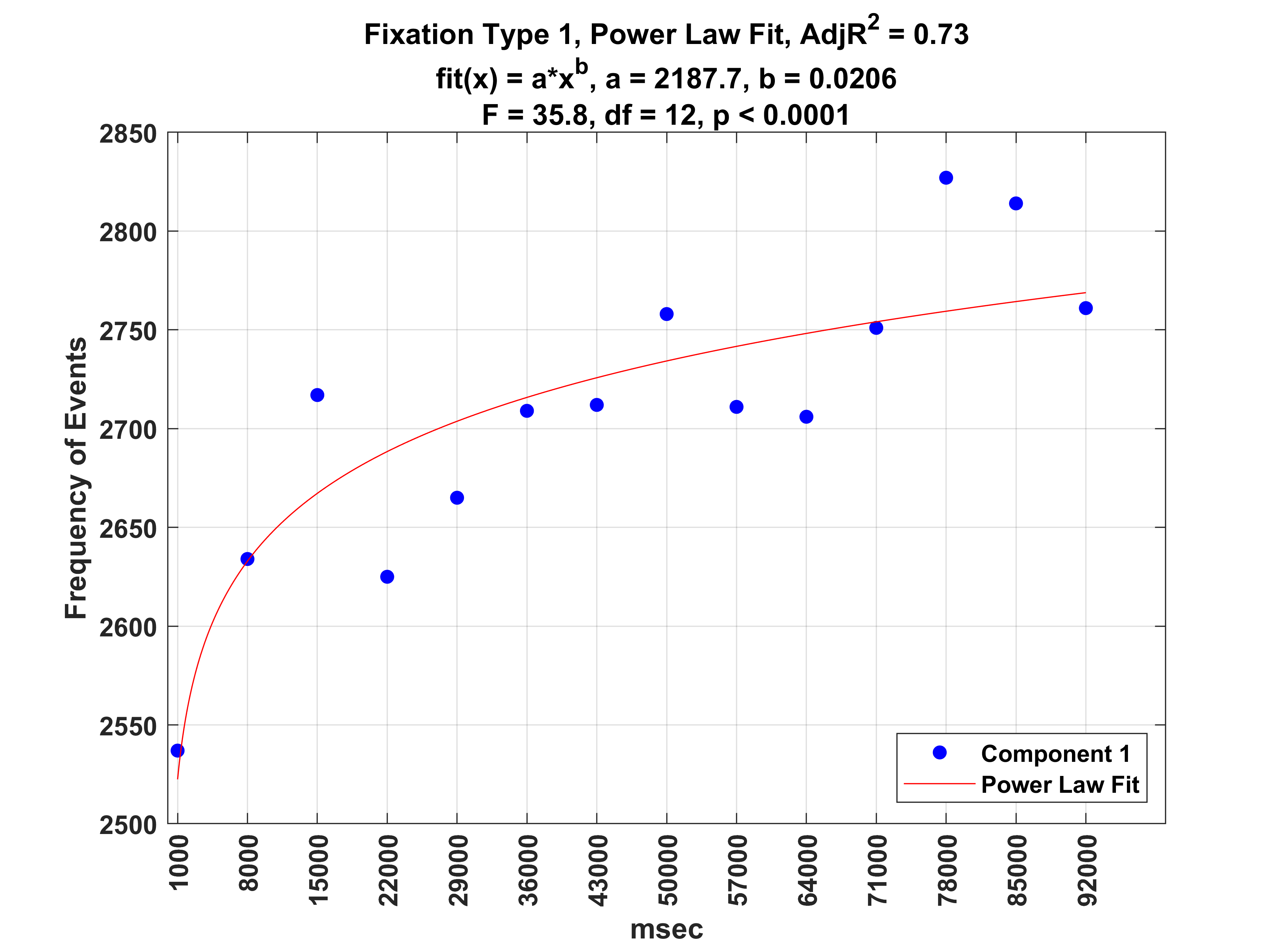}
\caption{{\bf Plot of the frequency of fixation type 1 over time.}}
\label{fig13}
\end{figure}

\begin{figure}[!ht]
\includegraphics[width=1.0\textwidth]{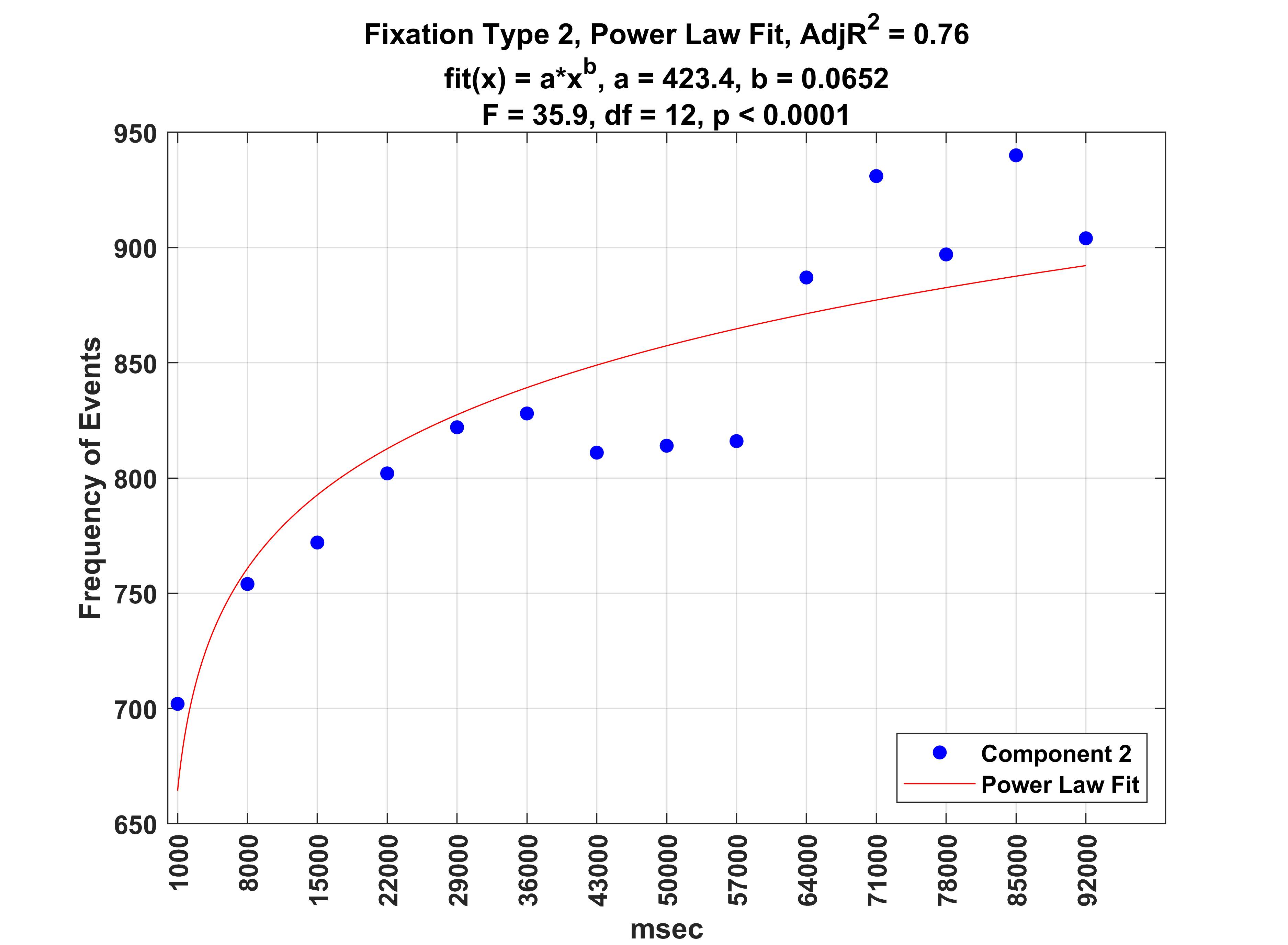}
\caption{{\bf Plot of the frequency of fixation type 2 over time.}}
\label{fig14}
\end{figure}

\begin{figure}[!ht]
\includegraphics[width=1.0\textwidth]{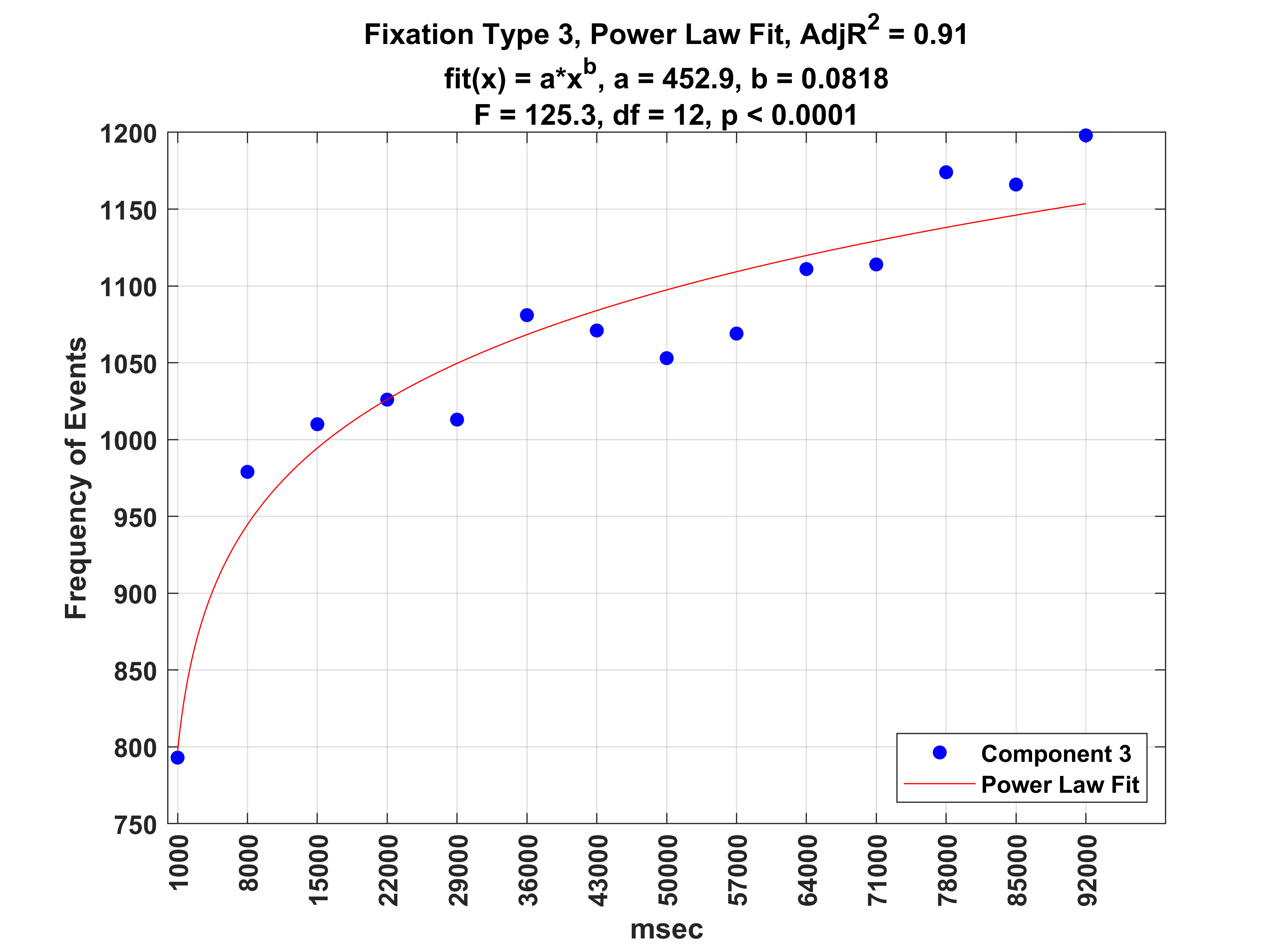}
\caption{{\bf Plot of the frequency of fixation type 3 over time.}}
\label{fig15}
\end{figure}

\begin{figure}[!ht]
\includegraphics[width=1.0\textwidth]{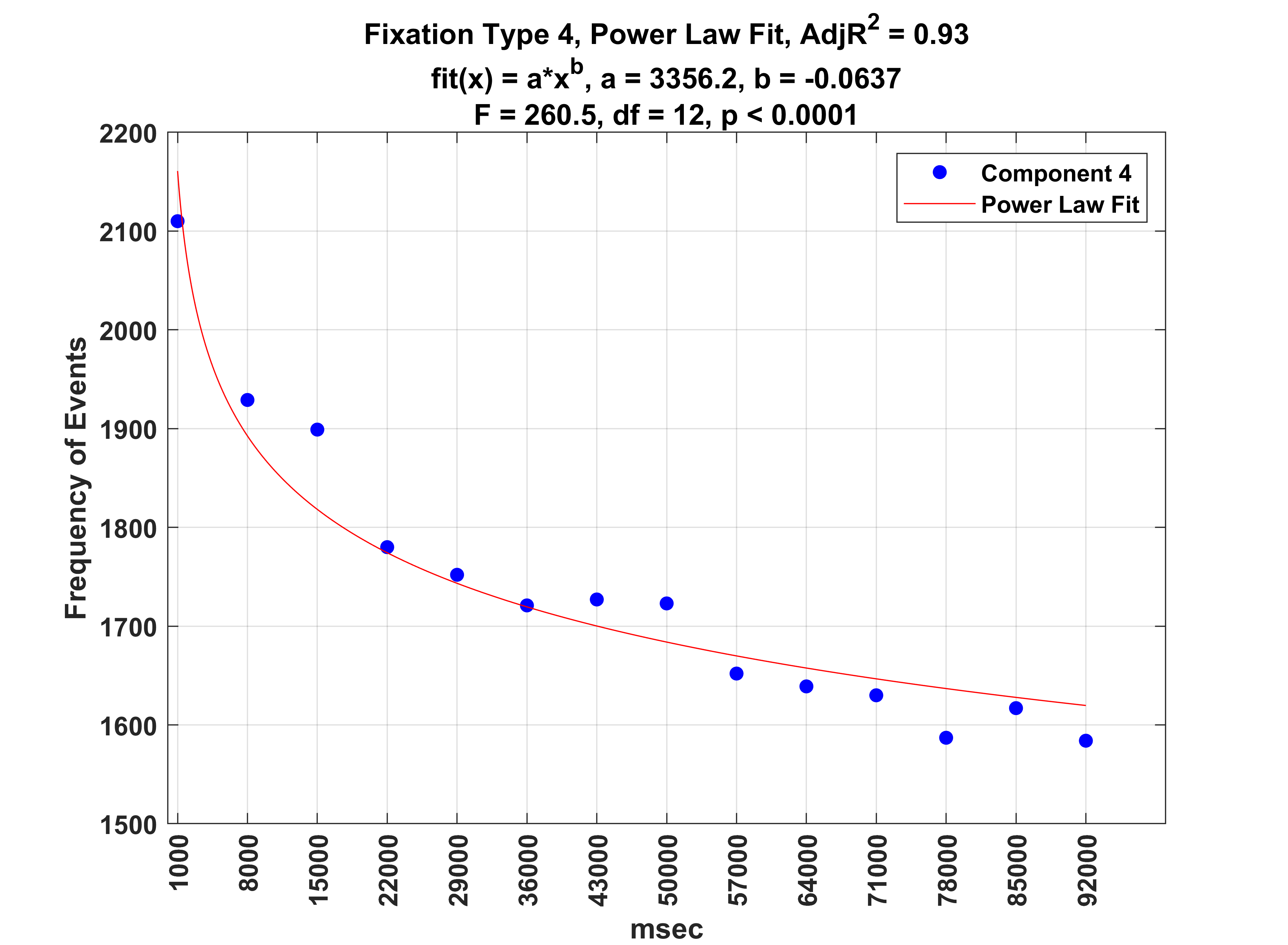}
\caption{{\bf Plot of the frequency of fixation type 4 over time.}}
\label{fig16}
\end{figure}

\begin{figure}[!ht]
\includegraphics[width=1.0\textwidth]{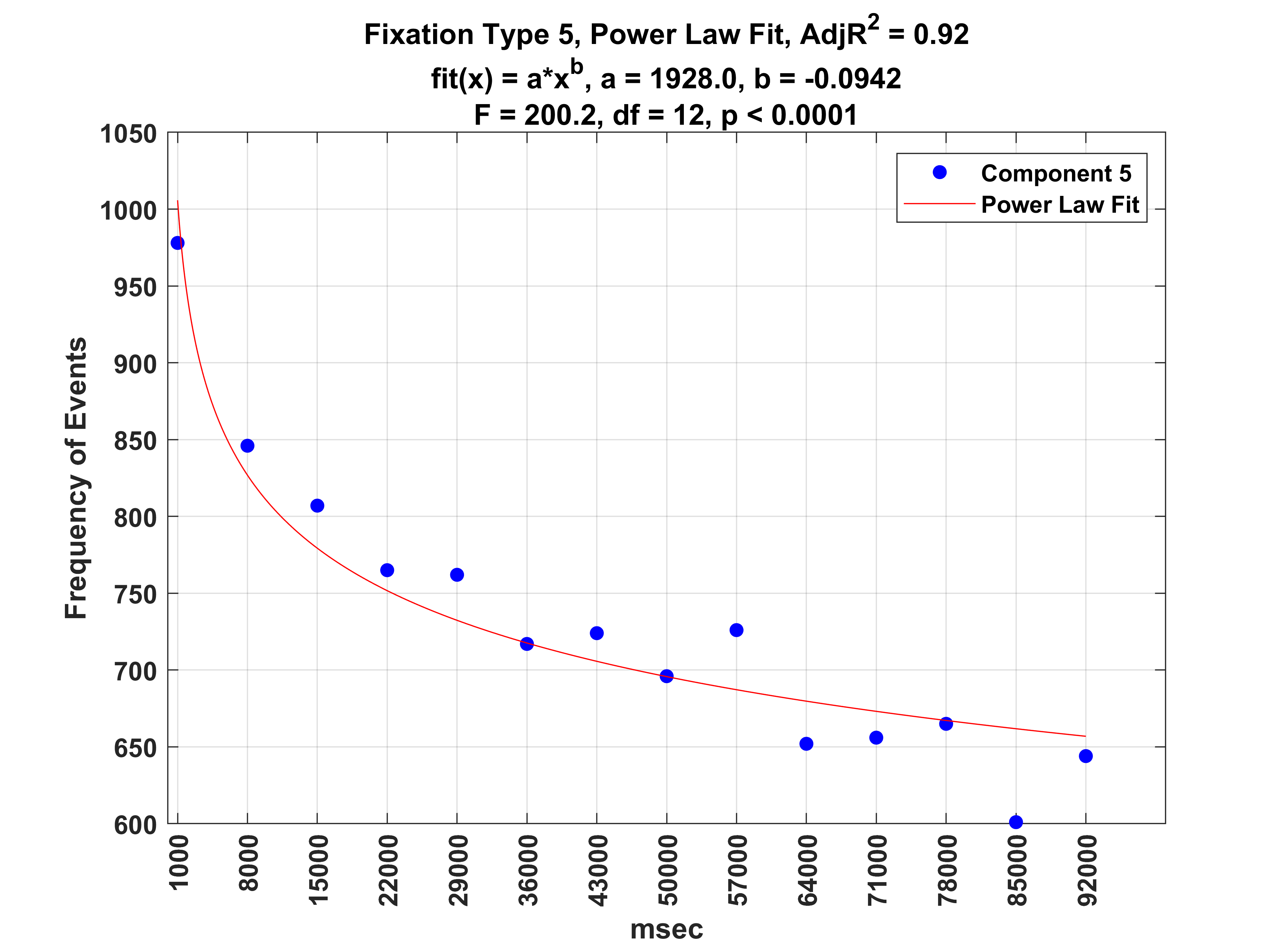}
\caption{{\bf Plot of the frequency of fixation type 5 over time.}}
\label{fig17}
\end{figure}

Figs. \ref{fig13},\ref{fig14}, \ref{fig15}, \ref{fig16} and \ref{fig17} illustrate the frequency of each fixation type by time period. For each fixation type, there were statistically significant changes in the frequency of events over these time periods (all $~p-values~<0.0001$).  Power law fits accounted for a large amount of variance in these estimates (from 78 to 93\% of the variance).  The first three (shortest) fixation types occurred more and more frequently over time (Figs. \ref{fig13}, \ref{fig14} and \ref{fig15}), whereas the last two (longest) fixation types (Figs. \ref{fig16} and \ref{fig17}) occurred less and less frequently over time.

\subsection*{Summary of Fixation Type Characteristics}
The key characteristics of the five Fixation Types are summarized in Table 3.  The percent of all fixations is based on the component weights from Table 2.


\begin{table*}
\caption{\textbf{Summary of characteristics of the five Fixation Types.}}
\begin{tabular}{|c|c|c|c|c|c|}
\thickhline
  & Mean & \% of all & \% of Time Fixation & \% of Time Fixation  & Frequency as \\
  & Length & Fixations & Type Follows & followed by a & a Function of\\
  & (ms) \footnote{Based on membership probability.} &  \footnote{Based on component weights from Table 2.} & Target Step & Corrective Movement & Time on Task\\ \hline
1 & 117 & 35 & 57.4 & 80.5 & Increase \\  \hline
2 & 240 & 15 & 10.0 & 61.4 & Increase\\  \hline
3 & 515 & 19 &  4.8 & 25.4 & Increase\\  \hline
4 & 776 & 22 &  4.3 & \footnote{Too few events to measure} & Decrease\\  \hline
5 & 939 & 10 & 23.4 & \textsuperscript{\textit{c}}  & Decrease \\  \hline
\end{tabular}
\end{table*}

\subsection*{Saccade accuracy over time period}
Saccade error increased as a function of TOT. See Fig \ref{fig18} for analysis of median saccade error over time and see Fig. \ref{fig19} for analysis of median saccade percent error (divided by total target movement) over time.  Although these relationships are statistically significant, the effects are small.  Mean saccade error starts at 1.18 dva at time block 1 and ends at 1.36 dva at the last time block.  Changes in percent saccade error start new 10\% and end near 11|\%.

\setcounter{figure}{17}

\begin{figure}[htbp]
\includegraphics[width=1.0\textwidth]{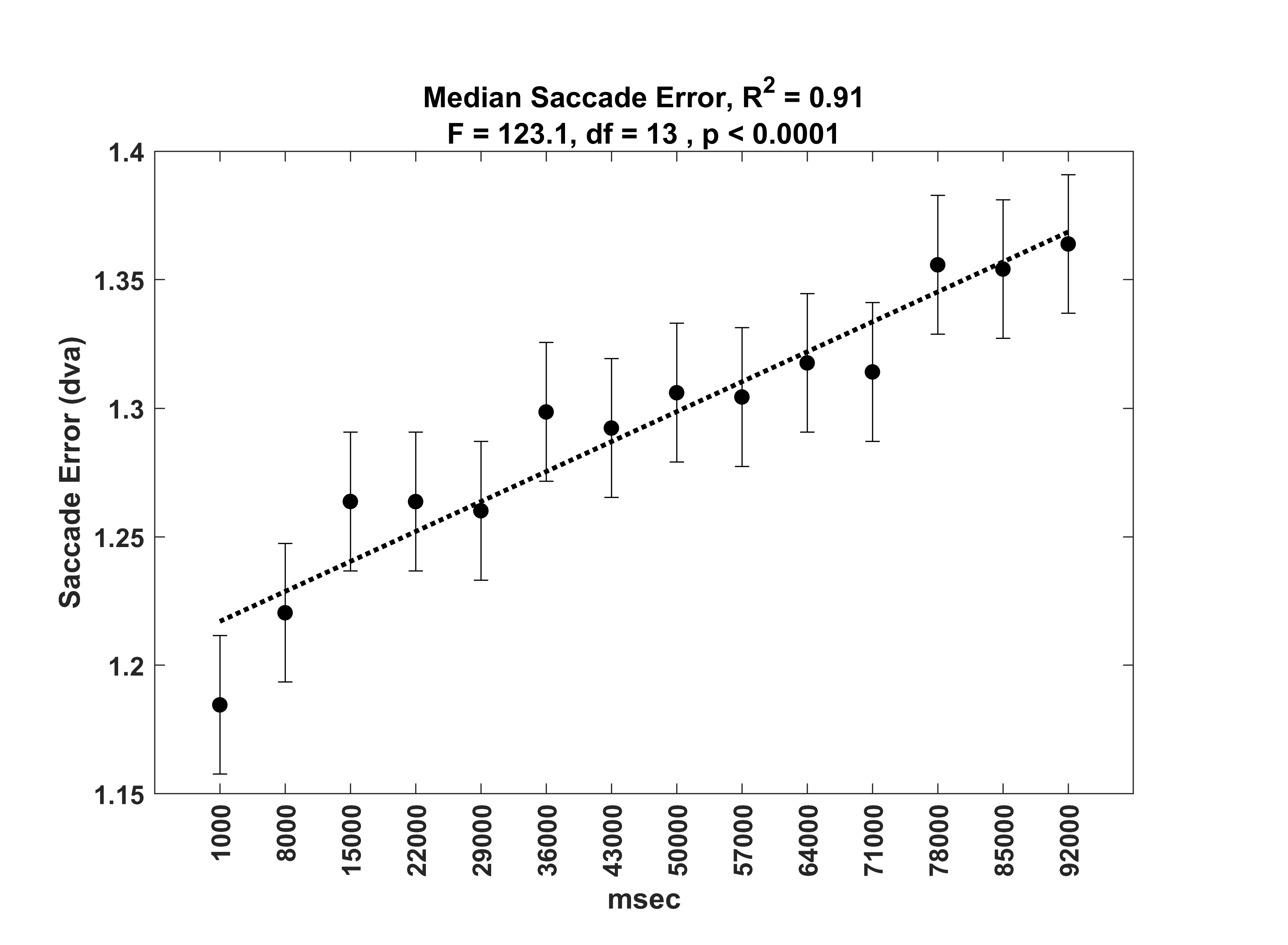}
\caption{{\bf Plot of the median saccade error over time.}}
\label{fig18}
\end{figure}

\begin{figure}[htbp]
\includegraphics[width=1.0\textwidth]{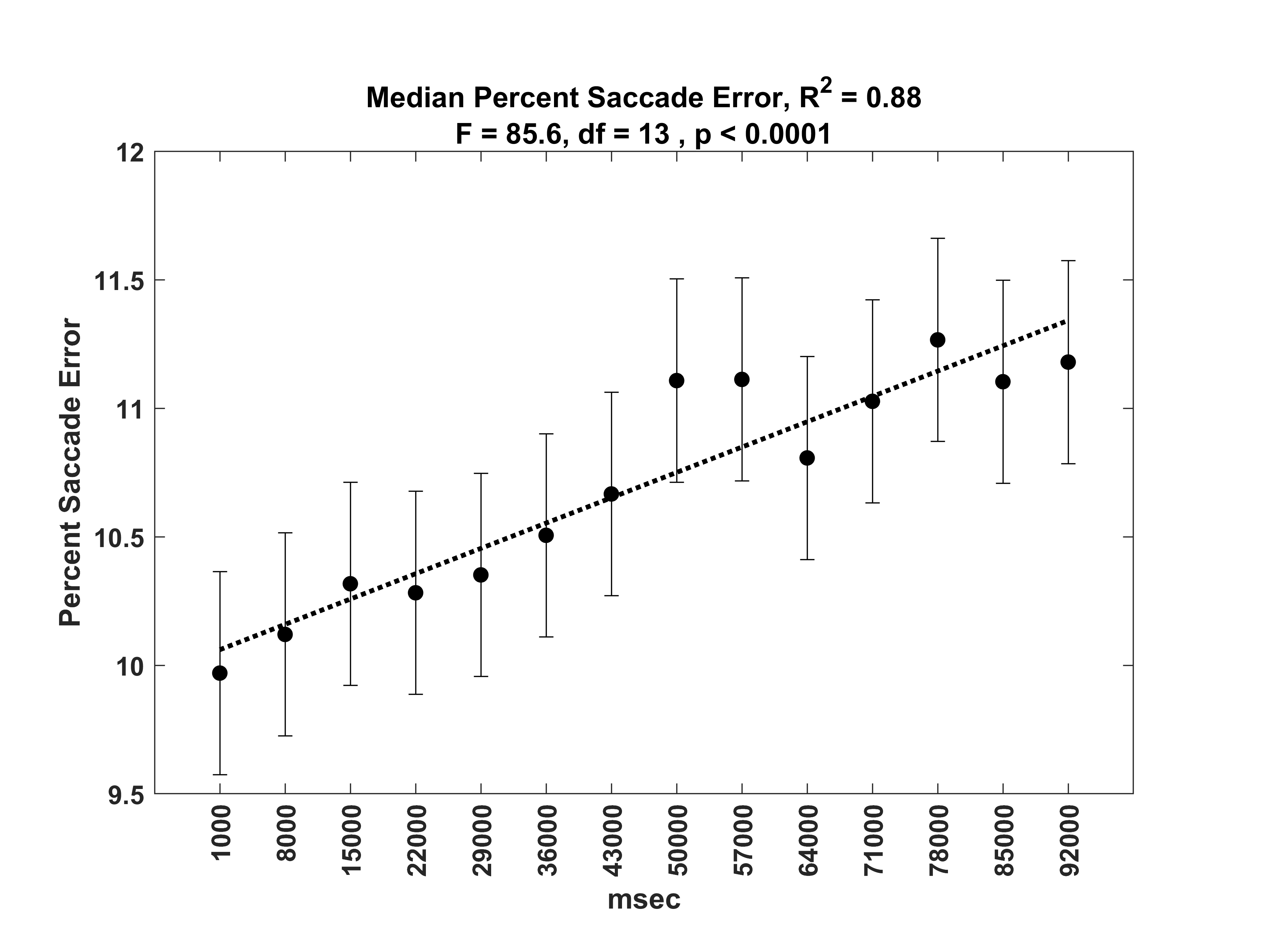}
\caption{{\bf Plot of percent saccade error over time.}}
\label{fig19}
\end{figure}

\section*{Discussion}
We evaluated eye-movement performance during a saccade task where the target jumped every 1 sec to a random location across the screen.  The task lasted 100 seconds (100 target steps).  Eye-movements were classified into fixations, saccades, post-saccadic oscillation and various types of artifacts.  We applied a Gaussian mixture model to the frequency histogram of fixation durations and found evidence for five fixation types.  The ``ideal'' response to such a stimulus would be an accurate saccade to the target after a typical saccade latency.  The eyes would remain on the target until the next target movement, 1 sec later.  Fixations resembling such ideal responses did occur, but they only followed a target step $\approx$ 23\% of the time (10\% of all fixations).  The median saccade latency for these ideal responses was 266 msec, and the median fixation duration for these fixations was $\approx$ 940 msec.  We labelled these fixations as Fixation Type 5.  More frequently ($\approx$ 57\% of the time), a target step was followed by a much shorter fixation ($\approx$ 117 msec, Fixation Type 1).  The saccade latency to these fixations was similar to the ideal response (272 msec).  Fixation Type 1 events were very likely to be followed by a corrective saccade.  
 
Fixations Type 2, 3 and 4 were much less likely to follow a target step (10\%, 4.8\% and 4.3\% of the time respectively).  Fixations of Type 2 ($\approx$ 211 msec) occurred over a wide range of latencies (median 332 msec) and, more often than not were also followed by a corrective saccade. Fixations of Type 3 were very heterogeneous with respect to their duration (range from 329 to 685 msec) and their time after target step (median $\approx$ 650 msec). They were rarely followed by a corrective saccade.  Fixations of Type 4 were similar to the ideal response (Fixation Type 5) in duration (mean 776 msec) but rarely followed a fixation step and occurred a median of 421 msec after target movement.

Total fixation duration across all fixation types declined markedly over the first 20 sec of the task and continued to decline more slowly through the remainder of the task.  A power law fit this decline in total fixation duration quite well with an $R^2$ of 0.94.  But this finding masks the effects of TOT on each fixation type.  The two longest fixation types (4 and 5) were relatively frequent at the start of the task but became sharply less frequent as the task continued.  On the other hand, the shorter fixation types (1, 2 and 3) became more and more frequent as the tasks continued.  All of these relationships were fit reasonably well with power law relationships.

We hypothesized that the ideal response might decline in frequency over time if saccade accuracy also declined with time.  We did find evidence for an increase in saccade error over time which might explain some of the decline in the frequency of the ideal response.  But the decrease in ideal fixation frequency events over time followed a power law and the increase in saccade error was linear with respect to time.  Also the magnitude of the increase in saccade error over time was not large.  Prior studies have found a relationship between fatigue and saccade accuracy.  For example, one night of sleep deprivation was associated with a decline in saccade accuracy in two studies \cite{SleepDep,SleepDep3}.  On the other hand, no change in saccade accuracy was reported after one night of sleep deprivation in another study \cite{SleepDep2}.  Also, although monkeys had slower saccades over time on task (500 saccades, $\approx$ one saccade per two seconds), accuracy was not diminished \cite{Monkey}.  Furthermore, Saito \cite{Saito} did not find a decline in saccade accuracy over a five hour eye-tracking study that subjects reported was quite fatiguing.

It is difficult to relate these findings to prior literature.  Our findings of 5 classes of fixation based on duration is heavily influenced by the unique structure of our task, with target steps every 1 sec.  Although our frequency histogram of fixation durations was fit very well with a Gaussian mixture model, one might question if fixation types 2 and 3 are truly Gaussian.  Perhaps fixation types 1 and 2 might be considered as a single positively skewed distribution, and fit with a log-normal or gamma or other distribution form. However, we are not aware of a robust method for fitting mixed distribution forms.  The differences we found across the five fixation types in terms of latency after the target step, presence of subsequent corrective saccade, and response to TOT support our analysis approach.

As noted in the introduction, most studies of oculomotor fatigue have evaluated changes over much longer intervals (minimum 18 min) than our task (100 seconds).  Furthermore, most of the changes we noted occurred over the first 20 seconds or so of our task.  We are unsure if the term ``fatigue'' is appropriately applied to such short time intervals and therefore we describe our effects as related to time on task (TOT).  Also as noted in the introduction, the findings on fixation duration and fatigue in the literature were mixed with no clear pattern emerging. 

\section*{Conclusion}
Based on a Gaussian mixture model we found evidence for five types of fixation based on fixation duration.  The two longest fixation types (greater than $\approx$ 650 msec) occurred less frequently over TOT whereas the three shorter fixations types occurred more frequently over TOT.  All of these temporal changes were well fit with power law functions.  These changes account for the decrease in total fixation duration we noted across our task.  
\section*{Acknowledgments}

%
%
%





\end{document}